\documentclass[english,onecolumn,draftclsnofoot,12pt]{IEEEtran}
\usepackage[latin9]{inputenc}
\usepackage{amsmath}
\usepackage{amsthm}
\usepackage{amssymb}
\usepackage{graphicx}

\makeatletter

\DeclareFontEncoding{LGR}{}{}
\DeclareRobustCommand{\greektext}{%
  \fontencoding{LGR}\selectfont\def\encodingdefault{LGR}}
\DeclareRobustCommand{\textgreek}[1]{\leavevmode{\greektext #1}}
\ProvideTextCommand{\~}{LGR}[1]{\char126#1}

\theoremstyle{plain}

\theoremstyle{remark}
\newtheorem{rem}{\protect\remarkname}

\IEEEoverridecommandlockouts
\usepackage{cite}
\usepackage{amsfonts}\usepackage{algorithmic}
\usepackage{textcomp}
\usepackage{xcolor}
\def\BibTeX{{\rm B\kern-.05em{\sc i\kern-.025em b}\kern-.08em
    T\kern-.1667em\lower.7ex\hbox{E}\kern-.125emX}}

\makeatother

\providecommand{\remarkname}{Remark}
\providecommand{\theoremname}{Theorem}

\begin{document}
\global\long\def\expect#1{\mathbb{E}\left[#1\right]}%

\global\long\def\abs#1{\left\lvert #1\right\lvert }%

\global\long\def\twonorm#1{\left\Vert #1\right\Vert }%

\global\long\def\brac#1{\left(#1\right)}%

\global\long\def\lgbrac#1{\log\left(#1\right)}%

\global\long\def\lnbrac#1{\ln\left(#1\right)}%

\global\long\def\lghbrac#1{\log\left(2\pi e\brac{#1}\right)}%

\global\long\def\cbrac#1{\left\{  #1\right\}  }%

\global\long\def\rline#1{\left.#1\right| }%

\global\long\def\sbrac#1{\left[#1\right] }%

\global\long\def\Det#1{\left|#1\right|}%

\global\long\def\prob#1{\mathbb{P}\brac{#1}}%

\global\long\def\eqdof{\doteq}%

\global\long\def\leqdof{\overset{.}{\leq}}%

\global\long\def\geqdof{\overset{.}{\geq}}%

\global\long\def\union{\bigcup}%

\global\long\def\inter{\bigcap}%

\global\long\def\real{\mathbb{R}}%

\global\long\def\tran{\mathsf{Tran}}%

\global\long\def\idty{\mathbbm{1}}%

\global\long\def\snr{\mathsf{SNR}}%

\global\long\def\inr{\mathsf{INR}}%

\title{Explicit Calibration of mmWave Phased Arrays with Phase Dependent
Errors\\
}
\author{\IEEEauthorblockN{Joyson Sebastian\IEEEauthorrefmark{1}, Pranav Dayal\IEEEauthorrefmark{2},
Walid AliAhmad\IEEEauthorrefmark{3} and Kee-Bong Song\IEEEauthorrefmark{4} } \IEEEauthorblockA{\textit{SOC R\&D Lab, Samsung Semiconductor, Inc.} \\
San Diego, USA \\
 Email: \IEEEauthorrefmark{1}joyson.vs@samsung.com, \IEEEauthorrefmark{2}pranav.dayal@samsung.com,
\IEEEauthorrefmark{3}walid.aa@samsung.com, \IEEEauthorrefmark{4}keebong.s@samsung.com}}
\maketitle
\begin{abstract}
We consider an error model for phased array with gain errors and phase
errors, with errors dependent on the phase applied and the antenna
index. Under this model, we propose an algorithm for measuring the
errors by selectively turning on the antennas at specific phases and
measuring the transmitted power. In our algorithm, the antennas are
turned on individually and then pairwise for the measurements, and
rotation of the phased array is not required. We give numerical results
to measure the accuracy of the algorithm as a function of the signal-to-noise
ratio in the measurement setup. We also compare the performance of
our algorithm with the traditional rotating electric vector (REV)
method and observe the superiority of our algorithm. Simulations also
demonstrate an improvement in the coverage on comparing the cumulative
distribution function (CDF) of equivalent isotropically radiated power
(EIRP) before and after calibration.
\end{abstract}


\section{Introduction}

In the 5G mmWave cellular systems, phased arrays are used for enhanced
coverage and directivity. This is achieved by beamforming, where phase
codes applied to the antennas help focus the antenna radiation in
the desired direction. We consider phased array with gain and phase
errors dependent on the phase applied. Our error model is motivated
from different error sources in a typical mmWave phased array solution.
Different routing lengths from the antennas to the phased array chip
cause phase errors. Each array chain has phase shifter connected to
gain stages which give rise to gain errors due to variation over process.
The phase shifters itself will have gain and phase errors depending
on the phase being applied.

We propose an explicit calibration method in presence of these errors.
The process of explicit calibration learns the phase and gain errors
in the phased array; by knowing these errors the codebook for the
phases applied on the antennas can be designed so that the beams point
in the intended directions and a good coverage pattern can be achieved.
The 3GPP FR2 documents provide requirements for minimum peak equivalent
isotropically radiated power (EIRP) and minimum EIRP at given percentile
of the cumulative distribution function (CDF) of the EIRP pattern.
The codewords need to be chosen carefully to meet the requirements
and our calibration assists in this process. The existing methods
in literature mostly focus on errors dependent only on the antenna
and not on the phase applied. The rotating-element electric field
vector (REV) method \cite{mano1982methodREVoriginal,REV_explained}
rotates the phase of one of the antennas while keeping the phase of
all other antennas at the default level. Then, the maximum and minimum
power under this rotation is recorded together with the phase applied
for maximum power. Using these, the relative phase and the relative
amplitude of the rotated antenna are obtained with respect to the
sum of vectors from all antennas. In \cite{takahashi2008fast}, a
modification of the REV method was proposed, where the phases of multiple
antenna elements were rotated simultaneously. These above methods
from literature require only power measurements for calibration.

There have been other works which need both power and phase measurements.
The phase toggle method was proposed in \cite{lee1993built}, where
the phase of one antenna is \textquoteleft toggled\textquoteright{}
by 180 degrees, while keeping the phase of all other antennas at the
default level. From the measurements made before and after 180 degree
rotation, the phase and amplitude of the toggled antenna can be obtained.
The Multi-Element Phase-toggle (MEP) method was proposed in \cite{hampson1999fast},
here multiple antenna elements are phase-rotated simultaneously and
the complex amplitudes of the antennas are obtained using a Fourier
transform. 

We propose a calibration method by explicitly calculating the phase
and gain errors on the antennas with the errors dependent on the phase
applied on the antenna. The gain errors are measured by turning ON
the antennas individually and measuring the power for each of its
phase. We assume for our system that only power measurements can be
made, hence the phase errors cannot be obtained from individual power
measurements. With pairwise measurements, with two antennas turned
ON, we can obtain the difference of the phases of the two antennas
(appearing in a cosine term), after the individual power measurements
are made. We obtain the phase errors of the antennas for each of the
phase computed with respect to the first phase of the first antenna,
i.e. the first phase of the first antenna is set as a reference. Our
measurements are made in such a way that the pairwise measurements
cover all the antennas and phases and the phase errors can be solved
uniquely from the system of equations arising from the measurements.
The solution of equations from the measurements are calculated assuming
that the measurements are noise-free. Once we obtain a solution for
the equations, we subsequently perform a least squared optimization
to minimize the error due to the noise in the measurements. We use
numerical simulations to evaluate the performance of our algorithm.
To compare with the REV method, we consider a case where the errors
are dependent only on the antennas and not on the phase applied. Our
numerical simulations show that our method performs better than the
REV method. We also illustrate a scenario of codebook design where
our calibration provides improvement in the coverage.

This report is organized as follows. In Section \ref{sec:System-Model-and},
we describe the system model and notations. In Section \ref{sec:Algorithm-for-measuring},
we describe our algorithm and optimization for measuring the errors.
In Section \ref{sec:Simulation-Results}, we give the simulation results
for our method. In Section \ref{sec:Conclusions}, we give the conclusions. 

\section{System Model and Notations\label{sec:System-Model-and}}

\begin{figure}[tbph]
\includegraphics[width=0.75\columnwidth]{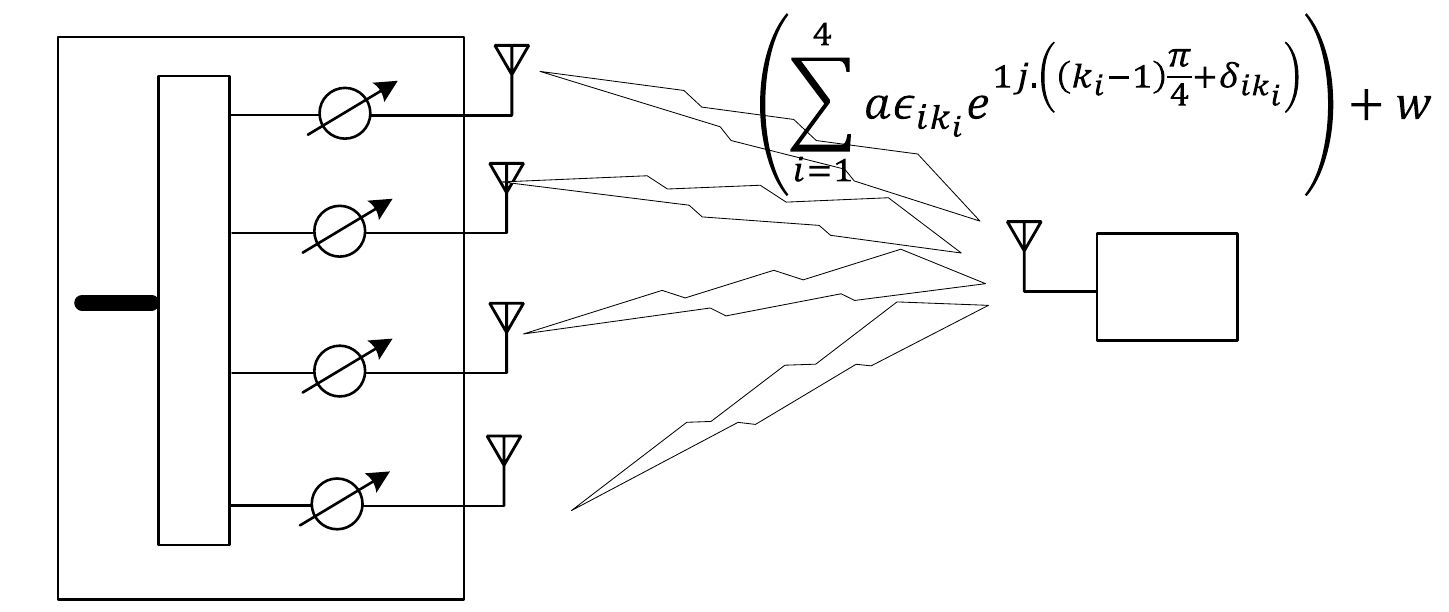}\caption{Illustration of measurement with all antennas turned ON}
\label{fig:power_mea}
\end{figure}

We consider an antenna array with $N$ elements. The phase level is
quantized with $Q$ bits. Our algorithm is described for the case
$N=4$ and $Q=3$, but it can be extended to the general case. For
our specific scenario, the possible phase outputs (without error)
are ${0,\pi/4,\ldots,7\pi/4}$. Including the errors, if all the antennas
are turned on, with $i^{\text{th}}$ antenna having phase index $k_{i}$,
the received electric field in the far-field in the boresight direction
is 
\begin{equation}
r=\brac{\sum_{i=1}^{4}a\epsilon_{ik_{i}}e^{1j.\brac{\brac{k_{i}-1}\frac{\pi}{4}+\delta{}_{ik_{i}}}}}+w
\end{equation}
where $a$ is the default amplitude of the signal from a single antenna
at receiver including the transmitter and receiver gain, and $w$
is an additive noise term. The term $\epsilon_{ik}$ denotes nonideal
amplitude scaling factor for $i^{\text{th}}$ antenna with $k^{\text{th}}$
phase, this arises due to variation in phase shifter amplitude gain,
amplifier gain and antenna gain. The term $\delta_{ik}$ denotes phase
error for $i^{\text{th}}$ antenna with $k^{\text{th}}$ phase, this
arises due to phase shifter error and phase mismatch between different
antenna array paths. The measurement of radiated power in the far-field
is illustrated in Fig.~\ref{fig:power_mea}.

\begin{figure}[tbph]
\includegraphics[width=0.6\columnwidth]{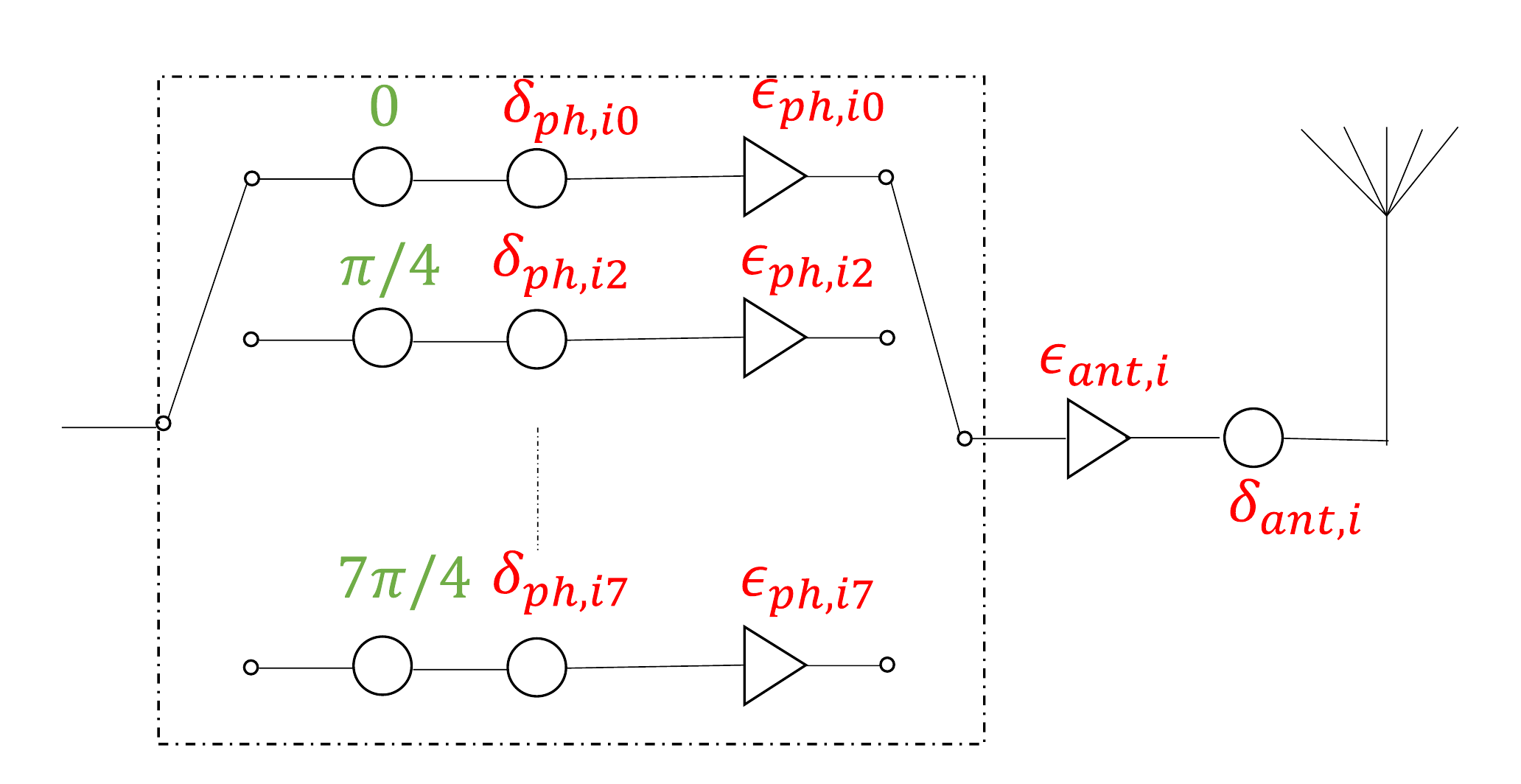}\caption{Illustration of errors for $i^{\text{th}}$ antenna.}
\label{fig:Error_model}
\end{figure}

We illustrate the errors of $i^{\text{th}}$ antenna in Fig.~\ref{fig:Error_model}.
Here $\epsilon_{\text{ph},ik}$ denotes the nonideal scaling factor
in phase shifter amplitude for $i^{\text{th}}$ antenna with $k^{\text{th}}$
phase, $\epsilon_{\text{ant},i}$ denotes the nonideal scaling factor
in the combination of amplifier gain and antenna gain for antenna
path $i$, and $\epsilon_{ik}=\brac{\epsilon_{\text{ph},ik}}\brac{\epsilon_{\text{ant},i}}$.
The phase error in the phase shifter is noted as $\delta_{\text{ph},ik}$,
while the phase error in the antenna path with respect to a nominal
phase common to all antenna paths is denoted by $\delta_{\text{ant},i}$,
and total phase error is noted as $\delta_{ik}=\delta_{\text{ph},ik}+\delta_{\text{ant},i}$.

For explicit calibration we would like to estimate $a\epsilon_{ik},\brac{k-1}\pi/4+\delta{}_{ik}$
for all $i,k$. We also use a shorter notation $\phi_{ik}=\brac{k-1}\pi/4+\delta{}_{ik}$,
also its estimate is indicated with a hat as $\hat{\phi}_{ik}$ .
We also define $b_{ik}=a\epsilon_{ik}$ and learn $b_{ik}$. Its estimate
is indicated as $\hat{b}_{ik}$. We assume that we can make only power
measurements and the measurements are made in the far-field boresight
direction. Note that the absolute value of $\phi_{ik}$ cannot be
measured using power measurements, since all power measurements remain
invariant under a constant added to $\phi_{ik}$, i.e if the system
had ${\phi_{ik}+c}$ for a constant c, all the power measurements
would remain the same. Hence we set $\phi_{00}$ as a reference for
measuring the phases. Hence the goal of our explicit calibration is
to measure $b_{ik}$ for all $(i,k)$ and $\phi_{ik}-\phi_{00}$ for
all ($i,k)\neq(0,0)$. We also use the notation Ant$i$Ph$k$ for
$\phi_{ik}$ for easily recognizing the index for antenna and phase.
The term $M_{ik}$ is the power measured with only $i^{\text{th}}$
antenna turned on at $k^{\text{th}}$ phase, and $M_{ik,mn}$ is the
power measured with $i^{\text{th}}$ antenna turned on at $k^{\text{th}}$
phase together with $m^{\text{th}}$ antenna turned on at $n^{\text{th}}$
phase.

\section{Algorithm for measuring the antenna errors\label{sec:Algorithm-for-measuring}}

\begin{figure}[tbph]
\includegraphics[width=3.25in]{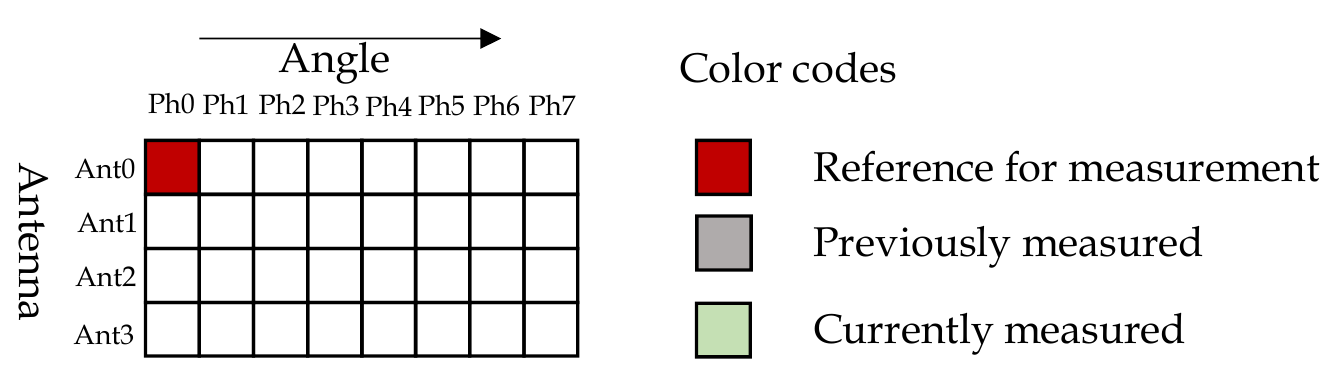}\caption{Illustration of antennas and phases.}
\label{fig:ant_phases}
\end{figure}

Our algorithm for measuring the errors are as follows:

\textbf{Step 1, Individual power measurements:} The first step of
our algorithm is to measure power with only single antenna turned
ON, going through all antennas and all phases. We denote this power
measurement by $M_{ik}$ where $i$ denotes the antenna index and
$k$ denotes the phase index. Thus this takes $4\times8=32$ measurements.
With these measurements, $\hat{b}_{ik}=\sqrt{M_{ik}}$ can be obtained. 
\begin{rem}
Since we can make only power measurements, the phase errors cannot
be measured from individual measurements. Using power measurements,
only the difference between the phases at two antennas are obtained,
for example with Ant0 and Ant1 ON with phase $\theta_{0}$ and rest
of the antennas OFF, the received power is $M_{10,00}=b_{00}^{2}+b_{10}^{2}+2b_{10}b_{00}\cos(\phi_{10}-\phi_{00})$
in the absence of thermal noise. We estimate $\cos(\hat{\phi}_{10}-\hat{\phi}_{00})$
as follows (in presence of noise):
\begin{equation}
\cos(\hat{\phi}_{10}-\hat{\phi}_{00})=\brac{4M_{10}M_{00}}^{-\frac{1}{2}}\cdot\brac{M_{10,00}-M_{10}-M_{00}}.
\end{equation}
In general $\cos(\phi_{ik}-\phi_{mn})$ can be obtained from the measurements
$M_{ik,mn},M_{ik},M_{mn}$ as 
\begin{equation}
\cos(\hat{\phi}_{ik}-\hat{\phi}_{mn})=\brac{4M_{ik}M_{mn}}^{-\frac{1}{2}}\cdot\brac{M_{ik,mn}-M_{ik}-M_{mn}}.\label{eq:cosine_measurement}
\end{equation}
\end{rem}
\textbf{Step 2, Setting the reference:} We choose $\phi_{00}$ as
reference and we can estimate other $\phi_{ik}$ with respect to $\phi_{00}$,
i.e., we can obtain $\hat{\phi}_{ik}-\hat{\phi}_{00}$. Henceforth
we will use $\hat{\phi}_{00}=0$ in this report. To describe our algorithm
for measuring the phase errors we use the illustration in Fig.~\ref{fig:ant_phases}
for antennas and phases.

\begin{figure}[tbph]
\includegraphics[width=0.5\columnwidth]{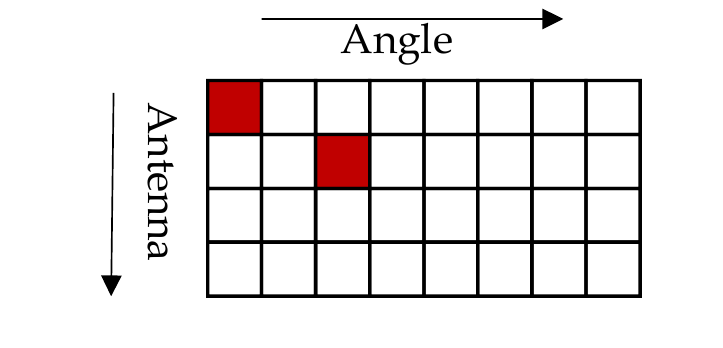}\caption{Two references for measuring the phases: Ant0Ph0 and Ant1Ph$R_{1}$}
\label{fig:two_refs}
\end{figure}

The measurements give only angle differences in the form $\cos(\hat{\phi}_{10}-\hat{\phi}_{00})$,
hence the phases cannot be obtained uniquely from a single measurement,
since $\cos^{-1}\brac{}$ is not unique in an interval of 2\textgreek{p},
i.e. $\cos\brac x=\cos(2\pi-x).$ Hence we need at least two measurements
with respect to two reference phases for uniquely obtaining the phases.
Fig.~\ref{fig:two_refs} illustrates the two references, the second
reference is chosen as Ant1Ph$R_{1}$ where Ph$R_{1}$ is to be determined,
$R_{1}$ could be any of $\cbrac{0,\ldots,7}.$ The second reference
could have been from any other antenna, but we choose Ant1. We will
describe in next step how to choose Ph$R_{1}$ and how to estimate
the phase of the second reference. 

\begin{figure}[tbph]
\includegraphics[width=3.25in]{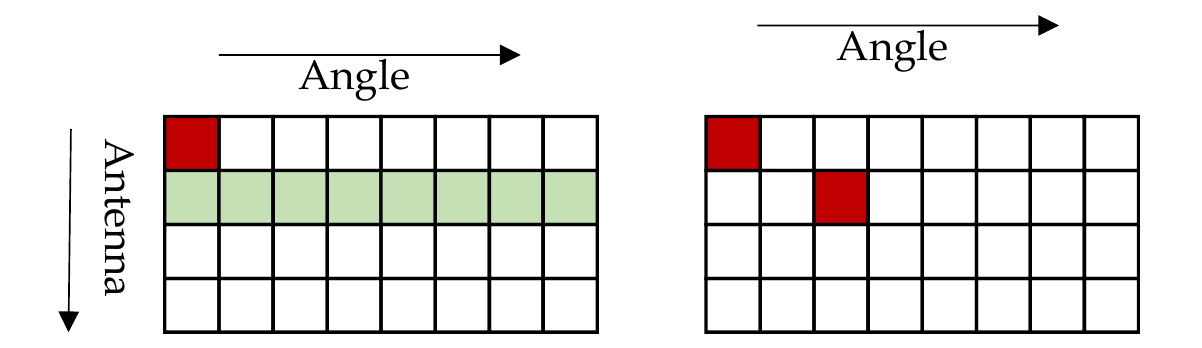}\caption{Figure 5: Choosing second reference.}
\label{fig:choosing_second_ref}
\end{figure}

\textbf{Step 3, Determining a second reference:} The other phases
$\phi_{ik},i>2$ are to be estimated with respect to both the references
to uniquely obtain $\hat{\phi}_{ik}$ from $\cos(\hat{\phi}_{ik})$
and $\cos(\hat{\phi}_{ik}-\hat{\phi}_{1R_{1}})$. For uniquely solving
$\hat{\phi}_{ik}$ from the values of $\cos(\hat{\phi}_{ik})$ and
$\cos(\hat{\phi}_{ik}-\hat{\phi}_{1R_{1}})$, we need that $\hat{\phi}_{1R_{1}}\neq0$.
Also we need $\hat{\phi}_{1R_{1}}\neq\pi$, this is because $\cos(x)=\cos(-x)$
and $\cos(x-\pi)=\cos(-x-\pi)$ which prevents a phase $x$ from being
resolved after obtaining $\cos()$ values with respect to 0 and \textgreek{p},
because --x also gives same $\cos\brac{}$ values with respect to
0 and \textgreek{p}. We propose to choose Ph$R_{1}$ so that $\abs{\hat{\phi}_{1R_{1}}}$
is close to \textgreek{p}/2. We estimate all the phases of Ant1 with
respect to Ant0Ph0 following \ref{eq:cosine_measurement} and choose
Ph$R_{1}$ so that $\abs{\cos(\hat{\phi}_{1R_{1}}-\hat{\phi}_{00})}=\abs{\cos(\hat{\phi}_{1R_{1}})}$
is closest to zero. This is illustrated in Fig.~\ref{fig:choosing_second_ref}
.

\begin{figure}[tbph]
\includegraphics[width=0.75\columnwidth]{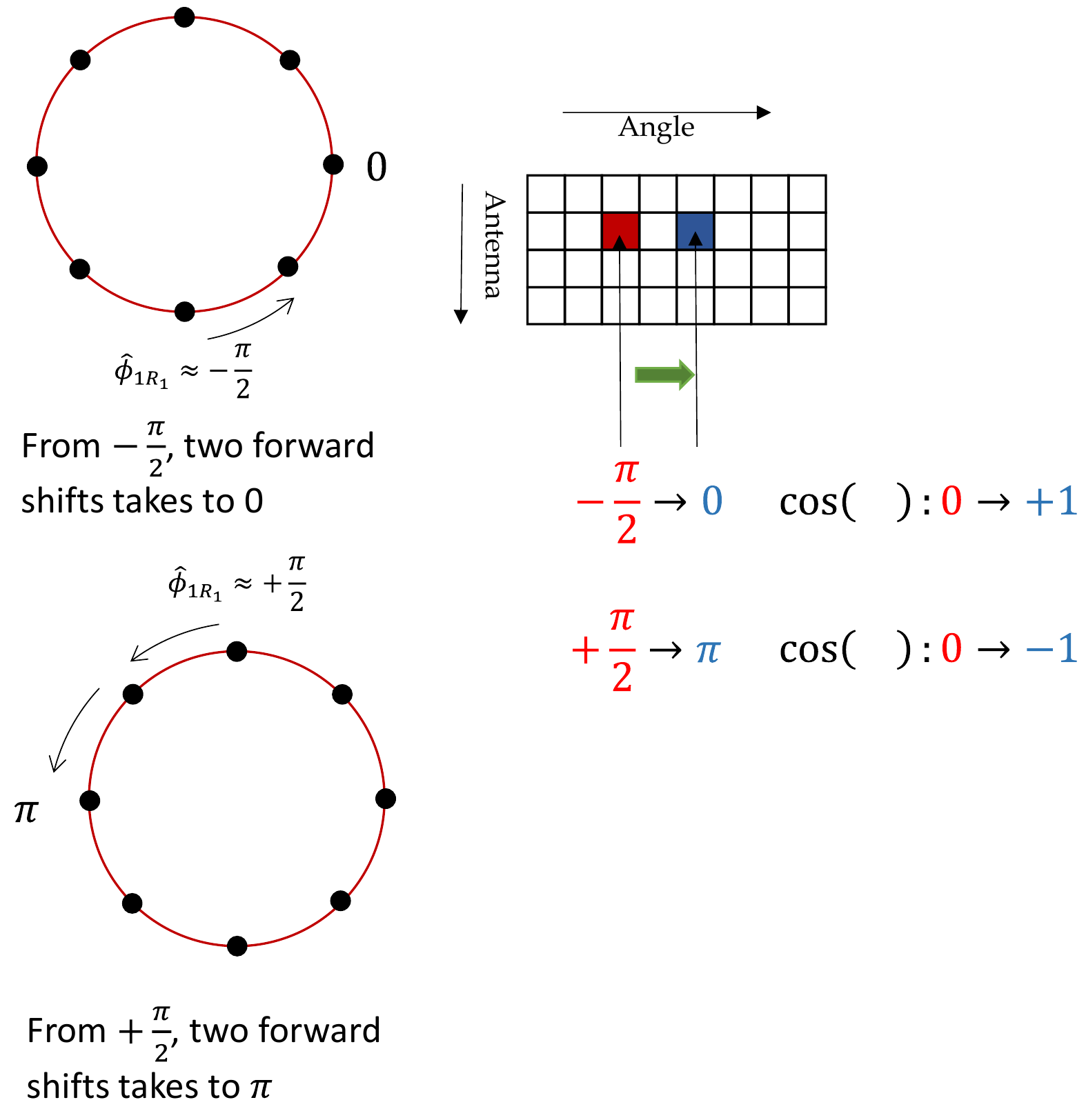}\caption{Resolving the sign of $\hat{\phi}_{1R_{1}}$ by looking at the value
of $\cos\protect\brac{\hat{\phi}_{1(R_{1}+2)\text{mod}8}}$ }
\label{fig:resolve_sign}
\end{figure}

To resolve the ambiguity of sign of $\hat{\phi}_{1R_{1}}$, we can
check the value of $\cos()$ values on the phase that is ahead of
$\hat{\phi}_{1R_{1}}$ by two indices. If $\hat{\phi}_{1R_{1}}$ was
close to $-\pi/2$, then shifting the phase value forward twice from
$R_{1}$ will cause the new phase to be close to zero, since the phase
quantization in our case is with separation $\pi/4$. Thus $\hat{\phi}_{1(R_{1}+2)\text{mod}8}$
will be close to 0, i.e., $\cos\brac{\hat{\phi}_{1(R_{1}+2)\text{mod}8}}$
will be close to +1. If $\hat{\phi}_{1R_{1}}$ was close to (+\textgreek{p})/2,
then shifting the phase value forward twice from $R_{1}$ will cause
the new phase to be close to \textgreek{p}, i.e., $\hat{\phi}_{1(R_{1}+2)\text{mod}8}$
will be close to \textgreek{p}, $\cos\brac{\hat{\phi}_{1(R_{1}+2)\text{mod}8}}$
will be close to -1. This is illustrated in Fig.~\ref{fig:resolve_sign}.

Thus if $\cos\brac{\hat{\phi}_{1(R_{1}+2)\text{mod}8}}$ is close
to +1, then we choose $\hat{\phi}_{1R_{1}}$ close to $-\pi/2$. If
$\cos\brac{\hat{\phi}_{1(R_{1}+2)\text{mod}8}}$ is close to -1 then
we choose $\hat{\phi}_{1R_{1}}$ close to $\pi/2$. This method for
resolving the sign of $\hat{\phi}_{1R_{1}}$ can be used for the case
with arbitrary number of quantization bits $Q$ for the number of
phases, with $Q\geq3$: we only need to look at whether $\cos\brac{\hat{\phi}_{1(R_{1}+2^{Q-1})\text{mod}2^{Q}}}$
is closer to +1 or -1.

\begin{figure}
\hfill{}\includegraphics[width=0.4\columnwidth]{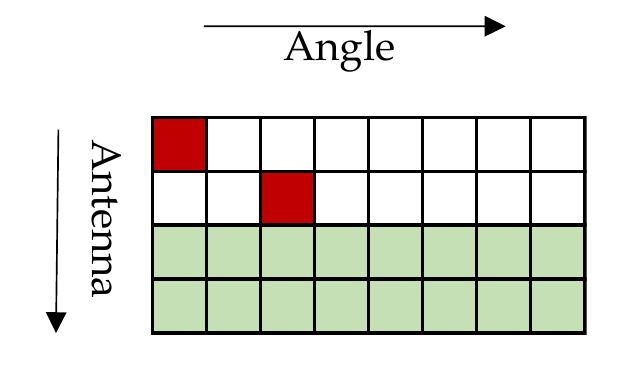}\hfill{}\includegraphics[width=0.4\columnwidth]{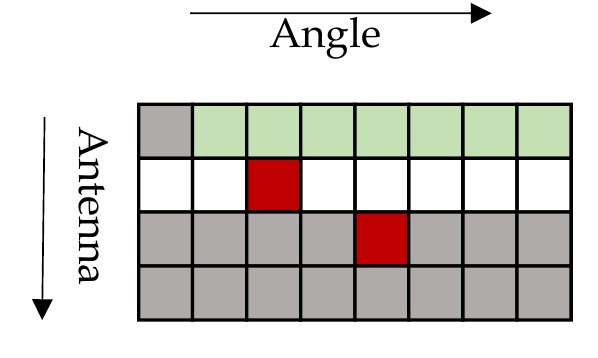}\hfill{}

\begin{minipage}[t]{0.5\columnwidth}%
\caption{\label{fig:Meas_Ant2_3}Estimation of phases on Ant2 and Ant3}
\end{minipage}%
\begin{minipage}[t]{0.5\columnwidth}%
\caption{\label{fig:Meas_Ant0}Estimation of phases on Ant0}
\end{minipage}\hfill{}
\end{figure}

\textbf{Step 4, Obtain all phases of Ant2, Ant3 with reference to
Ant0Ph0, Ant1Ph$R_{1}$:} Now the phases of the other antennas are
estimated with respect to the references. For example for estimating
$\phi_{20}$, Ant2 with Ph0 and Ant0 with Ph0 are turned ON with all
other antennas turned OFF and the power is measured. Then Ant2 with
Ph0 and Ant1 with Ph$R_{1}$ are turned ON with all other antennas
turned OFF and power is measured. From the two power measurements
$\cos(\hat{\phi}_{20})$and $\cos(\hat{\phi}_{20}-\hat{\phi}_{1R_{1}})$
are obtained. This step is illustrated in Fig.~\ref{fig:Meas_Ant2_3}.

We now illustrate how to estimate $\phi_{ik}$ from the two values
$A=\cos(\hat{\phi}_{ik}-\hat{\phi}_{00})$, $B=\cos(\hat{\phi}_{ik}-\hat{\phi}_{1R_{1}})$.
We include $\hat{\phi}_{00}$ in the calculations, to show the steps
with more generality, even though we had initially set $\hat{\phi}_{00}=0$.
\begin{equation}
A=\cos(\hat{\phi}_{ik}-\hat{\phi}_{00})=\cos\hat{\phi}_{ik}\cos\hat{\phi}_{00}+\sin\hat{\phi}_{ik}\sin\hat{\phi}_{00}
\end{equation}
\begin{equation}
B=\cos(\hat{\phi}_{ik}-\hat{\phi}_{1R_{1}})=\cos\hat{\phi}_{ik}\cos\hat{\phi}_{1R_{1}}+\sin\hat{\phi}_{ik}\sin\hat{\phi}_{1R_{1}}
\end{equation}
\begin{align}
\sbrac{\begin{array}{c}
A\\
B
\end{array}} & =\sbrac{\begin{array}{cc}
\cos\hat{\phi}_{00} & \sin\hat{\phi}_{00}\\
\cos\hat{\phi}_{1R_{1}} & \sin\hat{\phi}_{1R_{1}}
\end{array}}\sbrac{\begin{array}{c}
\cos\hat{\phi}_{ik}\\
\sin\hat{\phi}_{ik}
\end{array}}\nonumber \\
 & =K_{00,1R_{1}}\sbrac{\begin{array}{c}
\cos\hat{\phi}_{ik}\\
\sin\hat{\phi}_{ik}
\end{array}}
\end{align}
Now $K_{00,1R_{1}}$ is invertible if $\sin(\hat{\phi}_{00}-\hat{\phi}_{1R_{1}})\neq0$
i.e., if $\hat{\phi}_{1R_{1}}\neq\hat{\phi}_{00}+k\pi,k\in Z$. This
is ensured by our choice of $\hat{\phi}_{1R_{1}}$. By solving the
previous equation, we have
\begin{align}
\sbrac{\begin{array}{c}
\cos\hat{\phi}_{ik}\\
\sin\hat{\phi}_{ik}
\end{array}}=K_{00,1R_{1}}^{-1}\sbrac{\begin{array}{c}
A\\
B
\end{array}}
\end{align}

Now we can obtain $\hat{\phi}_{ik}=\text{Angle}(e^{1j.\hat{\phi}_{ik}}))=\text{Angle}(\cos\brac{\hat{\phi}_{ik}}+1j.\sin\brac{\hat{\phi}_{ik}})$.
When we have noisy measurements, we can say

\begin{equation}
\sbrac{\begin{array}{c}
u\\
v
\end{array}}=K_{00,1R_{1}}^{-1}\sbrac{\begin{array}{c}
A\\
B
\end{array}}
\end{equation}
\begin{equation}
\hat{\phi}_{ik}=\text{Angle}(u+1j.v)
\end{equation}

\textbf{Step 5, Determine Ant2PhR2:} For measuring all the phases
for Ant0, we need two references. One of the references can be Ant1Ph$R_{1}$.
We choose the other reference as Ant2Ph$R_{2}$. Similar to how Ant1Ph$R_{1}$
was chosen in relation to Ant0Ph0, requiring the two references to
be approximately \textgreek{p}/2 apart, we now choose Ant2Ph$R_{2}$
in relation to Ant1Ph$R_{1}$. Ph$R_{2}$ is chosen from the phases
of Ant2 such that so that $\cos(\hat{\phi}_{2R_{2}}-\hat{\phi}_{1R_{1}})$
is closest to zero. We do this by looking at the estimated values
of the phases of Ant2. 

\textbf{Step 6, Estimate all remaining phases of Ant0 with reference
to Ant1Ph$R_{1}$, Ant2PhR2:} Ant0Ph0 was set as zero for reference,
the remaining phases of Ant0 is estimated with reference to Ant1Ph$R_{1}$
and Ant2Ph$R_{2}$. This is illustrated in Fig.~\ref{fig:Meas_Ant0}
and is similar to Step 3.

\textbf{Step 7, Determine Ant2Ph$R_{3}$:} The remaining phases to
be estimated are of Ant1. For this we need two references, one of
them can be Ant0Ph0. The other reference can be from one of the remaining
antennas. We choose the other reference as Ant2Ph$R_{3}$, it is chosen
requiring the two references to be approximately \textgreek{p}/2 apart,
we make sure that $\abs{\cos(\hat{\phi}_{2R_{3}}-\hat{\phi}_{00})}=\abs{\cos(\hat{\phi}_{2R_{3}})}$
is closest to zero. We perform this by looking at the estimated values
of the phases of Ant2. 

\begin{figure}[tbph]
\includegraphics[width=0.4\columnwidth]{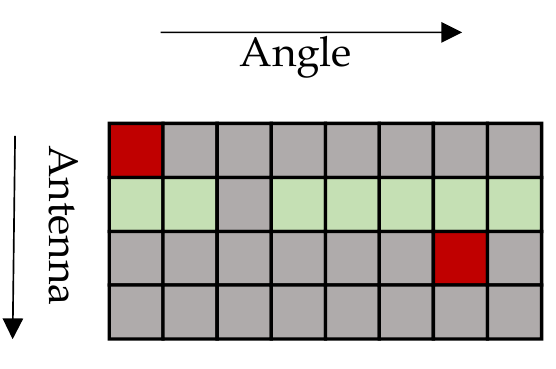}\caption{Phases of Ant1 is estimated with respect to Ant0Ph0 and Ant2Ph$R_{3}$}
\label{fig:meas_ant1}
\end{figure}

\textbf{Step 8, Estimate all remaining phases of Ant1 with reference
to Ant0Ph$R_{0}$, Ant2Ph$R_{3}$:} Now the remaining phases of Ant1
is estimated with reference to Ant0Ph0 and Ant2Ph$R_{3}$. This is
illustrated in Fig.~\ref{fig:meas_ant1}. Note that all the phases
of Ant1 were already measured with respect to Ant0Ph0 while Ant1Ph$R_{1}$
was being chosen. Now the new measurements are required only with
reference to Ant2Ph$R_{3}$.

The individual power measurements for obtaining the gain errors are
$4\times8=32$. For obtaining the phase, all the phases of all the
antennas require two measurements, except for Ant0Ph0 which needs
no measurement and for Ant1Ph$R_{1}$ which takes only one measurement;
this gives $2\times4\times8-2\times2+1=61$. Total number of measurements
is thus 93. For general N and Q the number of measurements can be
similarly calculated as $3\times N\times2^{Q}-3$

\subsection{Optimization }

We perform a least squared optimization to minimize the error in the
estimates due to noise in the measurements. Let $M_{ind}\brac l$
indicate the $l^{\text{th}}$ individual power measurement, let $i_{l}$
be the antenna index and let $k_{l}$ be the phase index involved
in $l^{\text{th}}$ measurement.
\begin{equation}
M_{ind}\brac l=\abs{b_{i_{l}k_{l}}e^{1j.\brac{\phi_{i_{l}k_{l}}}}+w_{l}}^{2}.\label{eq:ind_mea}
\end{equation}
Here $w_{l}$ is additive white Gaussian noise (AWGN). And let $M_{pair}(m)$
indicate the $m^{\text{th}}$ pairwise power measurement, $i_{m}^{(1)},i_{m}^{(2)}$
be the two antennas involved in $m^{\text{th}}$ pairwise power measurement.
Similarly $k_{m}^{(1)},k_{m}^{(2)}$ are the phase indices.
\begin{equation}
M_{pair}(m)=\abs{b_{i_{m}^{(1)}k_{m}^{(1)}}e^{1j.\phi_{i_{m}^{(1)}k_{m}^{(1)}}}+b_{i_{m}^{(2)}k_{m}^{(2)}}e^{1j.\phi_{i_{m}^{(2)}k_{m}^{(2)}}}+w_{m}}^{2}\label{eq:pair_wise_mea}
\end{equation}
Again $w_{m}$ is AWGN. We now change the variables to $a_{ik}e^{1j.\phi_{ik}}=a_{Rik}+1j.a_{Cik}$
and perform a least squares error optimization with respect to the
power measurements. 

\begin{align}
 & \underset{}{\text{Minimize}}\sum_{l}\brac{\abs{b_{Ri_{l}k_{l}}+1j.b_{Ci_{l}k_{l}}+w_{l}}^{2}-M_{ind}(l)}\nonumber \\
 & \hphantom{\underset{}{\text{Minimize}}}+\sum_{m}\Big(\left|b_{Ri_{m}^{(1)}k_{m}^{(1)}}+1j.b_{Ci_{m}^{(1)}k_{m}^{(1)}}+b_{Ri_{m}^{(2)}k_{m}^{(2)}}\right.\nonumber \\
 & \hphantom{\underset{}{\text{Minimize}}+\sum_{m}\Big(}\left.+1j.b_{Ci_{m}^{(2)}k_{m}^{(2)}}+w_{m}\right|^{2}-M_{pair}(m)\Big)
\end{align}
 The start point for the optimization is taken from the solution obtained
from our previous calculation. If we choose the error-free phase and
gain values as the starting point, we observed that the optimization
does not converge to the correct solution in our simulations. 

\section{Simulation Results\label{sec:Simulation-Results}}

\begin{figure}[tbph]
\includegraphics[width=0.75\columnwidth]{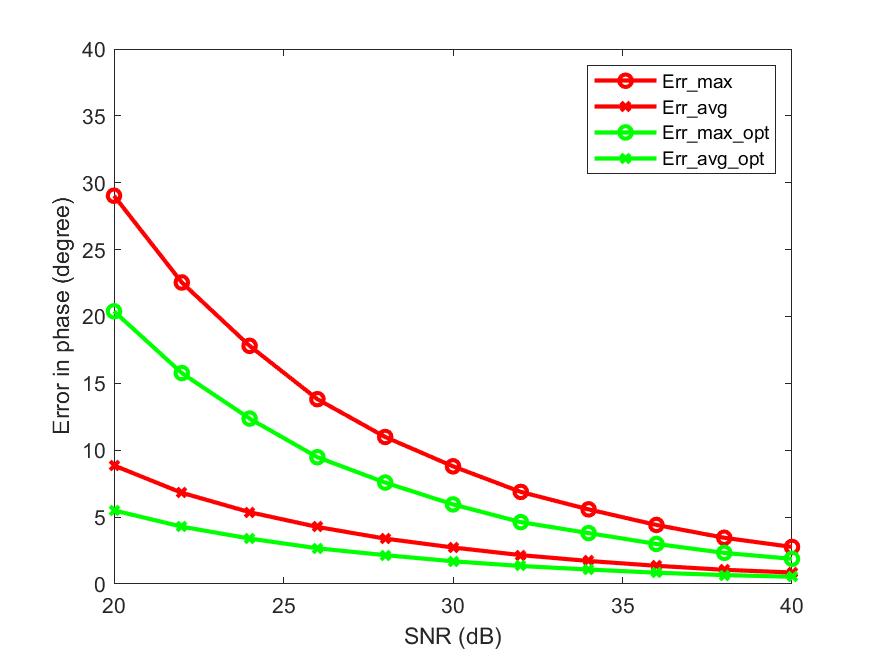}\caption{Comparison of error in phase calibration}
\label{fig:ph_error_opt_nonopt}
\end{figure}

\begin{figure}[tbph]
\includegraphics[width=0.75\columnwidth]{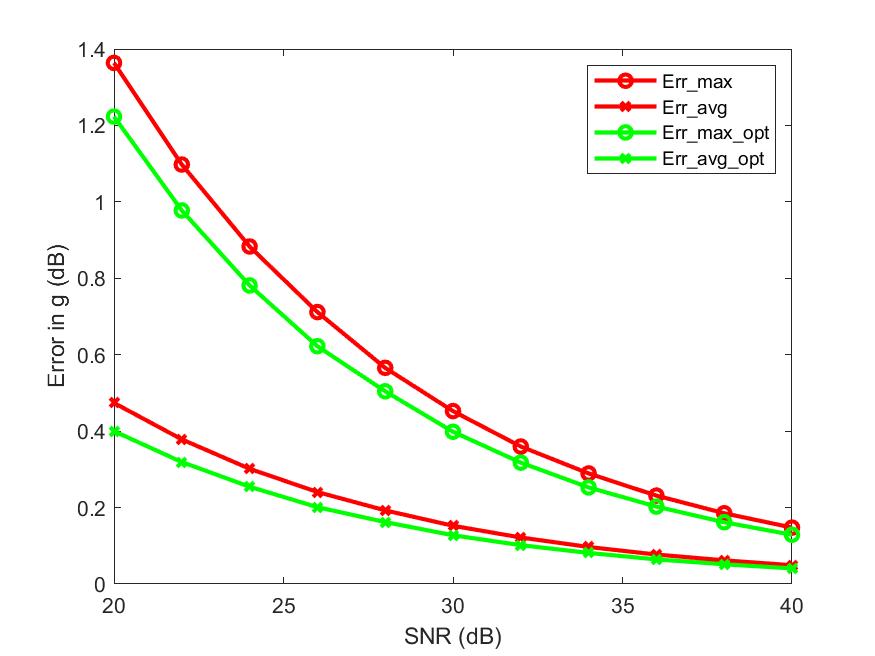}\caption{Comparison of percentage error in gain calibration.}
\label{fig:g_error_opt_nonopt}
\end{figure}

We simulate our algorithms with $b_{ik}$ drawn uniformly from $[-1.5,+1.5]$
in dB. The signal to noise ratio (SNR) in our measurements is determined
by the power of the noise $w$ in \eqref{eq:ind_mea},\eqref{eq:pair_wise_mea}
with signal power set at 0 dB. Phase error $\delta_{\text{ph},ik}$
is drawn uniformly from $[-10,+10]$ in degrees. Fixed phase errors
$\delta_{\text{ant},i}$ on antennas are drawn randomly from $[-180,+180]$
in degrees. For the solution obtained from our algorithm without the
least squared optimization, we use Err\_max to denote the maximum
error among all the phase estimates for a given array, and we have
Err\_avg as the average error. Similarly, after the least squared
optimization, we have the terms Err\_max\_opt and Err\_avg\_opt. The
error in calibration of the phase is given in Fig.~\ref{fig:ph_error_opt_nonopt}
and the error in calibration of the amplitude gain is given in Fig.~\ref{fig:g_error_opt_nonopt}.
The plots are obtained by averaging over 1000 iterations for each
SNR. The average of the maximum error and average error for phases
are calculated as $\expect{\underset{i,k}{\text{max}}\brac{\underset{l\in\mathbb{Z}}{\text{min}}\abs{\phi_{ik}-\hat{\phi}_{ik}+l2\pi}}}$
and $\expect{\underset{l\in\mathbb{Z}}{\text{min}}\abs{\phi_{ik}-\hat{\phi}_{ik}+l2\pi}}$
respectively where the expectation is over the iterations and $(i,k)\neq(0,0)$.
Similarly the average of maximum error for gain is calculated in dB
as $\expect{\underset{i,k}{\text{max}}\brac{20\log10\abs{\abs{b_{ik}-\hat{b}_{ik}}/b_{ik}+1}}}$,
and the average error for gain is calculated as $\expect{20\log10\abs{\abs{b_{ik}-\hat{b}_{ik}}/b_{ik}+1}}$where
the expectation is over the iterations and all $i,k$.

\begin{figure}[tbph]
\includegraphics[width=0.75\columnwidth]{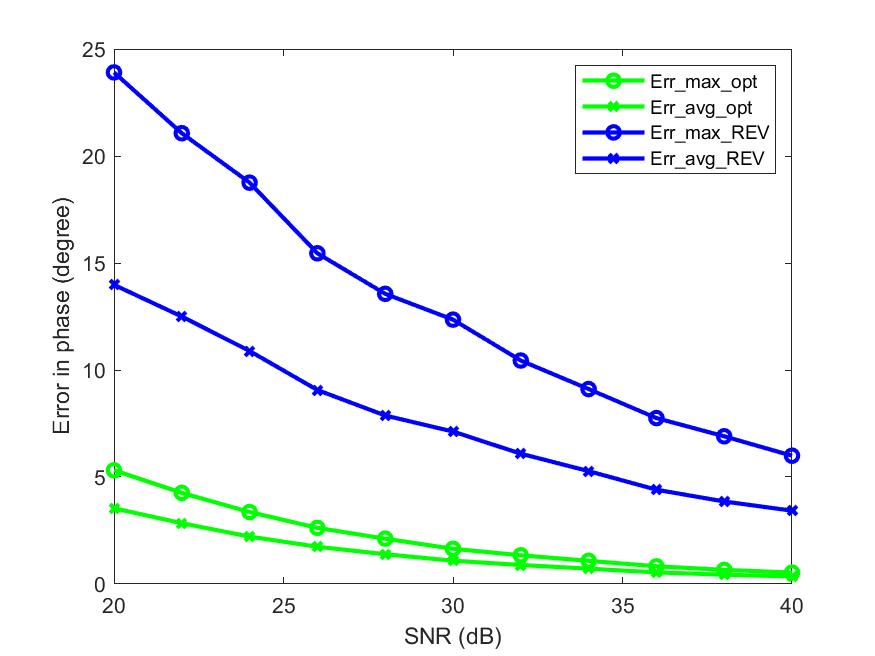}\caption{Comparison with REV method for error in phase calibration}
\label{fig:ph_error_REV_comp}
\end{figure}

\begin{figure}[tbph]
\includegraphics[width=0.75\columnwidth]{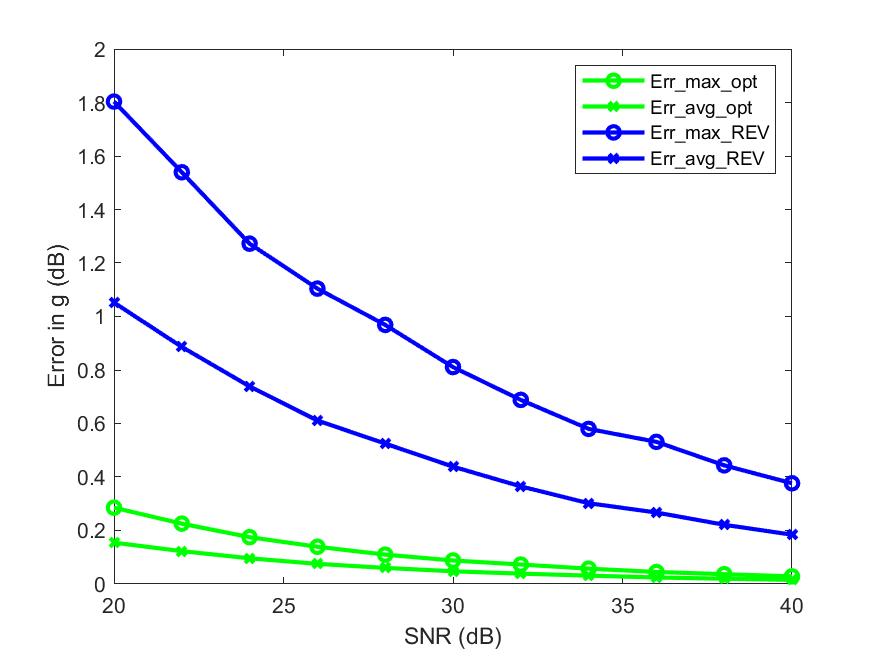}\caption{Comparison with REV method for percentage error in gain calibration.}
\label{fig:g_error_REV_comp}
\end{figure}

We simulate the REV method with $\delta_{\text{ant},i}$ drawn in
the same way as in the previous simulation. However, there are no
phase dependent errors $\delta_{\text{ph},ik}$ this case. Also, we
have $b_{i}$ dependent only on the antenna, instead of $b_{ik}$
dependent on both antenna and phase. For the solution from the REV
method, we denote the maximum error among the three $\delta_{\text{ant},i}$
as Err\_max\_REV and the average error over the three phases as Err\_avg\_REV.
For rotating the phase of each antenna in the REV methods, we go through
eight phases. Hence with two iterations of REV method and four antennas,
the total number of measurements is $2\times4\times8=64$. We compare
the performance of REV method with our algorithm with the least squared
optimization. Our original algorithm is designed for the case with
phase dependent errors. Hence for the case with no phase-dependent
errors, we average the errors across the phases. For our solution,
we denote the maximum error among the three $\delta_{\text{ant},i}$
as Err\_max\_opt and the average error over the three phases as Err\_avg\_opt.
The error in calibration of the phase is given in Fig.~\ref{fig:ph_error_REV_comp}
and the error in calibration of the amplitude gain is given in Fig.~\ref{fig:g_error_REV_comp}.
We note that our method performs better than the REV method. 

\begin{figure}[tbph]
\includegraphics[width=0.6\columnwidth]{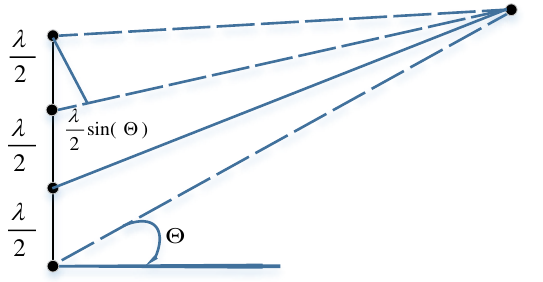}\caption{Antenna illustration.}
\label{fig:antenna_rays}
\end{figure}

We also demonstrate the usefulness of our method to improve the coverage.
We consider an ideal codebook designed for an omnidirectional linear
4-antenna array with separation between antennas as $\lambda/2$,
with $\lambda$ denoting the wavelength of the radio wave from the
antenna. This is illustrated in Fig.~\ref{fig:antenna_rays}. Due
to symmetry we can describe the antenna pattern using single angle
parameter $\Theta$. The path difference of $\brac{\lambda/2}\sin\brac{\Theta}$
corresponds to a phase difference of $\pi\sin\brac{\Theta}$. Similarly
considering all the rays from four antennas, and a code with phases
$[\phi_{0i_{0}},\phi_{1i_{1}},\phi_{2i_{2}},\phi_{3i_{3}}]$ on the
antennas, the power in the direction $\Theta$ is 
\begin{align}
P(\Theta)= & \left|1+e^{1j.\brac{\phi_{1i_{1}}+\pi\sin\brac{\Theta}}}\right.\nonumber \\
 & \left.+e^{1j.\brac{\phi_{2i_{2}}+2\pi\sin\brac{\Theta}}}+e^{1j.\brac{\phi_{3i_{3}}+3\pi\sin\brac{\Theta}}}\right|^{2}.\label{eq:eirp_actual}
\end{align}

For a quantized codebook with multiples of $\pi/4$ as the permitted
phases in the absence of errors, we design codes for directions\footnote{These directions were chosen by trial and error to ensure a good spherical
coverage.} $[90,19,40,-19,-40,0]$ in degree. For each direction we choose the
codeword that gives maximum power according to \eqref{eq:eirp_actual}.

\begin{figure}[tbph]
\includegraphics[width=0.75\columnwidth]{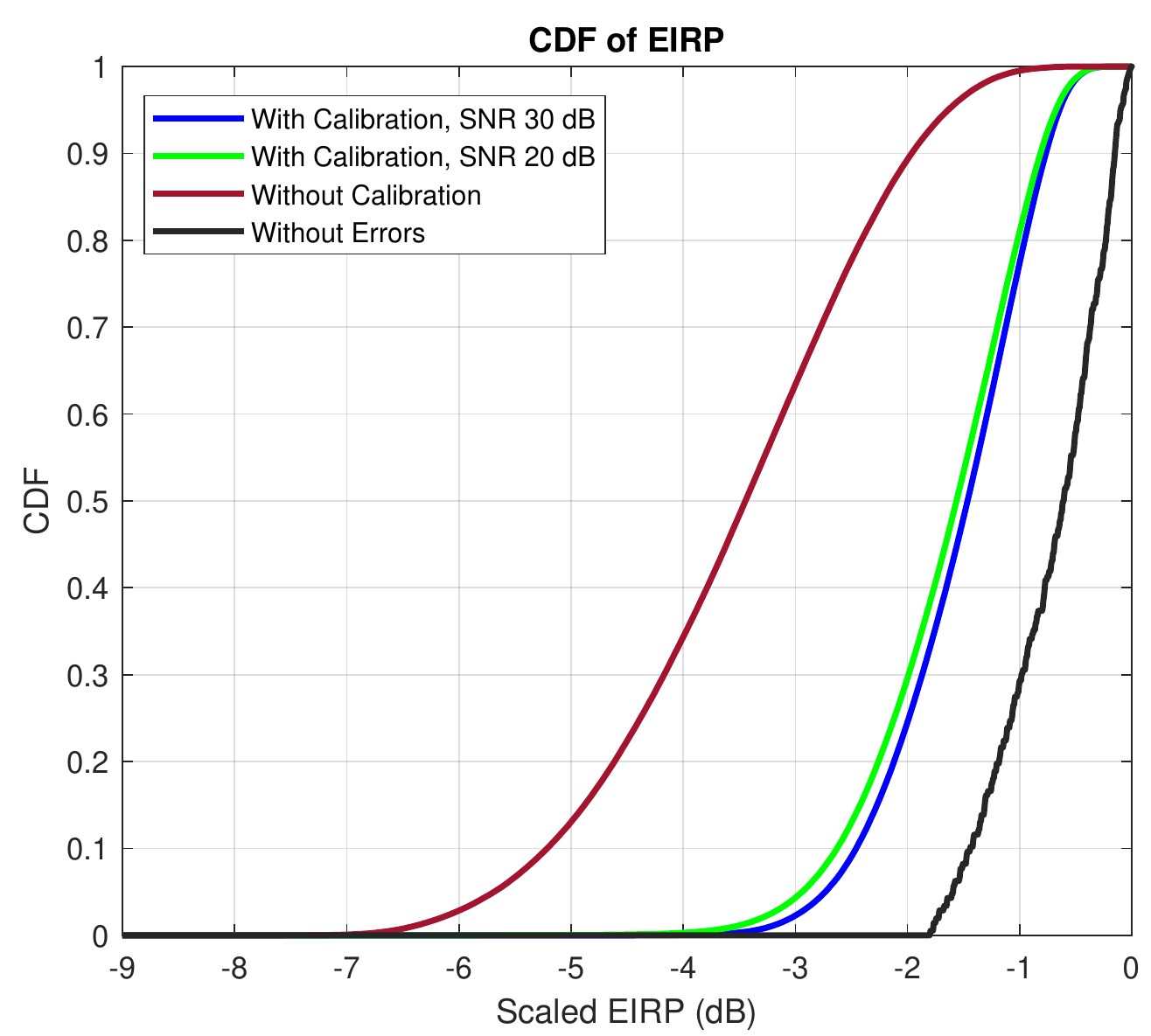}\caption{Comparison of CDF performance.}
\label{fig:cdf_comparison}
\end{figure}

Let $\hat{\phi}_{ik}$ be the estimated phases in presence of errors
at the antennas obtained by our algorithm. For a given code which
gives phases $[\hat{\phi}_{0i_{0}},\hat{\phi}_{1i_{1}},\hat{\phi}_{2i_{2}},\hat{\phi}_{3i_{3}}]$
on the antennas and corresponding magnitudes $[\hat{b}_{0i_{0}},\hat{b}_{1i_{1}},\hat{b}_{2i_{2}},\hat{b}_{3i_{3}}]$,
the power in a direction \textgreek{J} is \emph{calculated} as 
\begin{align}
 & \hat{P}(\Theta)\nonumber \\
 & =\left|\hat{b}_{0i_{0}}e^{1j.\hat{\phi}_{0i_{0}}}+\hat{b}_{1i_{1}}e^{1j.\brac{\hat{\phi}_{1i_{1}}+\pi\sin\brac{\Theta}}}\right.\nonumber \\
 & \quad\left.+\hat{b}_{2i_{2}}e^{1j.\brac{\hat{\phi}_{2i_{2}}+2\pi\sin\brac{\Theta}}}+\hat{b}_{3i_{3}}e^{1j.\brac{\hat{\phi}_{2i_{2}}+3\pi\sin\brac{\Theta}}}\right|^{2}.\label{eq:eirp_estimated}
\end{align}
Subsequently for each code direction, we can choose the code that
gives the maximum power $\hat{P}(\Theta)$ according to our calculation.
For the codeword designed for error-free scenario and for the codeword
designed with estimated errors, we then calculate the observed EIRP.
We obtain the CDF (averaged over multiple error instances) of scaled
EIRP in a sphere around the phased array. The CDF is illustrated in
Fig.~\ref{fig:cdf_comparison}. We consider two cases SNR of 20 dB
and SNR of 30 dB in measurements for our calibration algorithm. The
variables $b_{ik},\delta_{\text{ph},ik},\delta_{\text{ant},ik}$ are
generated as described at the beginning of this section. The EIRP
scaling is with respect to the maximum possible EIRP in the simulation
setup .i.e. when the $b_{ik}$'s are $1.5$ dB and combine coherently.
We observe that our calibration significantly improves the coverage.
In terms of 3GPP requirements, on the average, with 20 dB SNR we can
observe about 1.5 dB improvement in 50\%-tile EIRP and 0.73 dB improvement
in 99\%-tile EIRP compared to the case without calibration.

\section{Conclusions\label{sec:Conclusions}}

We proposed a method for explicit calibration of phased arrays using
only power measurements. We find the gain errors and phase errors
using $3\times N\times2^{Q}-3$ measurements where $N$ is the number
of antennas and $Q$ is the number of quantization bits for phased
array. Our method consists of an algorithm to solve for the values
of the errors by simple step by step calculation for the phase values
compared to chosen reference values. Subsequently we apply a least
squared optimization around the initial solution, to reduce the estimation
error. Our method uses relatively low number of measurements, demonstrates
more accurate results compared to the REV method, and provides improvement
in coverage.

\bibliographystyle{IEEEtran}
\bibliography{references}

\end{document}